 \documentclass[10pt]{aastex63}		
 \usepackage{natbib}
\def\gapprox{\lower.4ex\hbox{$\;\buildrel >\over{\scriptstyle\sim}\;$}}
\def\lapprox{\lower.4ex\hbox{$\;\buildrel <\over{\scriptstyle\sim}\;$}}

\shortauthors{Aschwanden}
\shorttitle{Universal Constants in SOC Systems}

\begin{document}
\renewcommand{\topfraction}{0.95}
\renewcommand{\bottomfraction}{0.95}
\renewcommand{\textfraction}{0.05}
\renewcommand{\floatpagefraction}{0.95}
\renewcommand{\dbltopfraction}{0.95}
\renewcommand{\dblfloatpagefraction}{0.95}


\title{Universal Constants 
	in Self-Organized Criticality Systems}
 
\author{Markus J. Aschwanden}
\affil{Lockheed Martin, Solar and Astrophysics Laboratory (LMSAL),
       Advanced Technology Center (ATC),
       A021S, Bldg.252, 3251 Hanover St.,
       Palo Alto, CA 94304, USA;
       e-mail: markus.josef.aschwanden@gmail.com}

\begin{abstract}
The occurrence frequency distributions of fluxes (F)
and fluences or energies (E) observed in the 
majority (in 18 out of 23 cases of astrophysical 
phenomena) are found to be consistent with 
the predictions of the fractal-diffusive self-organized 
criticality (FD-SOC) model, which predicts power law 
slopes with universal constants of 
$\alpha_F=(9/5)=1.80$ for the flux, and
$\alpha_E=(5/3)\approx 1.67$ for the fluence, 
respectively.  
The theoretial FD-SOC model is based on the
fractal dimension, the flux-volume proportionality, 
and classical diffusion. The universal  
scaling laws predict the size distributions 
of numerous astrophysical phenomena, such as 
solar flares, 
stellar flares,
coronal mass ejections (CME), 
auroras, 
blazars,
active galactic nuclei (AGN), 
black-hole systems (BH),
galactic fast radio bursts (FRB), 
gamma-ray bursts (GRB), and 
soft gamma-ray repeaters (SGB). 
In contrast we identify 5 outliers 
of astrophysical phenomena, including
coherent solar radio bursts,
random solar radio bursts,
solar energetic partices (SEP),
cosmic rays, and 
pulsar glitches, which are not
consistent with the standard FD-SOC model,
and thus require different physical mechanisms.
\end{abstract}

\section{	INTRODUCTION 				}  

Nonlinear physics operating in astrophysical systems that
are governed by {\sl self-organized criticality (SOC)}
(Bak et al.~1987), has been studied in over 6000 publications,
reaching out to almost every science discipline, such as
planetary physics, solar physics, stellar physics,
galactic physics, geophysics, biophysics, financial physics,
or sociophysics, which have been described in the textbooks of
Bak (1996), Jensen (1998), Ilachinski (2001), Hergarten (2002),
Sornette (2004), Scott (2007), 
Aschwanden (2011a, 2013a, 2013b, 2019a, 2025), 
Pruessner (2012), Galam (2012), Charbonneau (2017). Most recent 
reviews have been presented in Watkins et al.~(2016), Aschwanden 
et al.~(2016); Sharma et al.~(2016), and McAteer et al.~(2016). 

The concept of SOC has been evolved and expanded over time.
Today we can distinguish between two schools: (i) the microscopic
concept of cellular automaton algorithms (Bak et al.~1987), and 
(ii) the macroscopic
concept of physical scaling laws (Aschwanden 2014, 2015, 2022). 
Instead of using the next-neighbor interactions of cellular 
automata, we quantify the spatial inhomogeneity in terms of 
the fractal dimension $D_d$ for the Euclidean domains 
$d=1$ (lines), $d=2$ (areas), or $d=3$ (volumes). 
Each fractal domain has a maximum fractal dimension of $D_d=d$, 
a minimum value of $D_d=d-1$, and a mean value of $D_V=d-1/2$ 
(Fig.~1),
\begin{equation}
	D_V={(D_{\rm V,max}-D_{\rm V,min}) \over 2} = d-{1 \over 2}  \ .
\end{equation}
For most applications in the (observed) 3-D world, the dimensional
domain $d=3$ is appropriate, which implies a fractal dimension 
$D_V=2.5$. However, if 2-D areas are observed, the fractal 
dimension is $D_A=1.5$ and the dimensionality is $d=2$. 
In this work we will mostly make use of the 3-D fractal domain, 
while the 2-D domain is discussed elsewhere (e.g., Aschwanden 2022). 
The fractal volume $V$ is then defined by the standard
(Hausdorff) fractal dimension $D_V$ in 3-D and the length scale 
$L$ (Mandelbrot 1977),
\begin{equation}
	V(L) \propto L^{D_V} \ .
\end{equation}
We formulate the statistics of SOC
avalanches in terms of size distributions (or occurrence frequency
distributions) that obey the scale-free probability distribution
function (Fig.~2), (Aschwanden 2014, 2015, 2022), 
\begin{equation}
	N(L) dL \propto L^{-d} dL \ ,
\end{equation}
where $d=1,2,3$ represent the Euclidean dimensions 
of the fractal domains
and $L$ is the length scale of a SOC avalanche. From this
scale-free relationship, the power law slopes $\alpha_x$
of other SOC parameters $x=[A,V,F,E,T]$ can be derived, 
such as for the area $A$, 
the volume $V$, the flux $F$, the fluence or energy $E$, and 
the duration $T$. The resulting power law slopes $\alpha_x$ 
can then be obtained mathematically by the method of variable 
substitution $x(L)$, by inserting the inverse function
$L(x)$ and its derivative $|dL/dx|$, 
\begin{equation}
	N(x) dx = N[L(x)] \left| {dL \over dx} \right| dL 
	= \ x^{-\alpha_x} dx \ ,
\end{equation}
such as for the flux $x=F$,
\begin{equation}
	\alpha_F = 1 + {(d-1) \over D_V \gamma} = {9 \over 5} = 1.80 \ ,
\end{equation}
or for the fluence $x=E$,
\begin{equation}
	\alpha_E = 1 + {(d-1) \over d \gamma} = {5 \over 3} = 1.67 \ .
\end{equation}
where $\gamma$ is the nonlinearity coefficient in the flux-volume
relationship, 
\begin{equation}
	F \propto V^\gamma = \left( L^{D_V} \right)^\gamma \ ,
\end{equation}
which degenerates to proportionality $F \propto V$ for $\gamma=1$.
The proportionality between the (fractal) volume $V$ and the
(observed) flux $V$ is depicted in Fig.~(3). In astrophysical
high-temperature plasmas, the volume $V$ is approximately 
proportional to the number of electrons in a region of instability, 
while the flux $F$ is proportional to the number of emitting photons,
which implies that the photon-to-electron ratio is approximately
constant and justifies the assumption of the flux-volume 
proportionality, in the case of incoherent emission mechanisms.

While this brief derivation of
Eqs.~(1)-(7) expresses the two main assumptions of fractality 
and linear flux-volume relationship for $\gamma=1$,
a third assumption needs to be brought in that takes
the spatio-temporal evolution into account (e.g., 
see scale-free statistics of spatio-temporal auroral
emission, Uritsky et al.~2002), which can be 
accomplished by the assumption of (classical) 
diffusive transport,
\begin{equation}
	L \propto T^{\beta/2} = T^{1/2} \ ,
\end{equation}
with the transport coefficient $\beta=1$. We call this
model the {\sl standard fractal-diffusive self-organized
criticality (FD-SOC)} model, 
defined by [$d=3, \gamma=1, \beta=1$], while the
generalized FD-SOC model allows for variable coefficients
[$d, \gamma, \beta$]. Note that there is a subtle paradigm 
shift from microscopic to macroscopic concepts. 
The classical Bak-Tang-Wiesenfeld (1987)
model mimics transport in SOC avalanches by cellular
automaton redistribution rules in the microscopic
world, while it is conceptualized by diffusive transport
in models of SOC avalanches. We prefer the FD-SOC model in
quantitative modeling of SOC processes, since cellular
automaton algorithms require numerical methods and cannot
be expressed analytically as a function of time. 
However, the diffusive
transport in a SOC avalanche appears to generate 
similar avalanches as the next-neighbor interactions 
of cellular automata. 

In this study we are testing the universality of 
power law slopes for fluxes ($\alpha_F=1.80$) and
fluences ($\alpha_E=1.67$) for a set of 23 astrophysical
phenomena. We find that most of the relevant observations
are consistent with the standard FD-SOC model, but can
identify also cases that require non-SOC models. 
In Section 2 we describe the data analysis and results,
which include the selection of observations (Section 2.1),
the definition of flux and fluence (Section 2.2),
the nonparametric statistics of observed power law slopes (Section 2.3),
the results of power law slopes for fluxes (Section 2.4)
and fluences (Section 2.5), 
statistical outliers (Section 2.6),
the nonlinear scaling of SEP events (Section 2.7),
the effects of coherent and incoherent radiation (Section 2.8), 
the power laws of random dize distributions (Section 2.9), and  
the establishment of universal constants in terms of the FD-SOC model
(Section 2.10). 
	
\section{	DATA ANALYSIS AND RESULTS 		}  

\subsection{	Selection of Observational Data  	}

The input data used in this study covers a comprehensive
data set of published occurrence frequency distributions 
(aka size distributions), obtained from publications that
contain power law slopes of fluxes, ($\alpha_F$), and fluences
or energies, ($\alpha_E$), observed and measured in astrophysical 
phenomena. The selection procedure involves searching and 
identifying of relevant keywords in the titles and 
abstracts in publications, using the NASA Astronomical 
Database System (ADS).
A list of 23 astrophysical phenomena is compiled in Tables 1 and 2,
which span from solar distances to galactic scales. 
The third and fifth columns in Table 1 specify the means
and standard deviations of published power law slopes,
$\alpha_F$ and $\alpha_E$, listed separately for fluxes 
and fluences of astrophysical phenomena. An estimate of 
the number of analyzed size distributions can be obtained 
from the sum of the cases, which amounts to ($\sum N_F=170$) 
for fluxes, and ($\sum N_E=104$) for fluences or energy, as 
listed in Table 1 (second and forth columns).   

\subsection{	Definition of Flux and Fluence	}	

The flux is defined in physical units of [energy/time],
usually measured at the (background-corrected) peak flux
$F$ at the peak time $t_p$ of an event,
\begin{equation}
	F={\rm max} \left[ f(t=t_p) \right] - f_{\rm BG} \  ,
\end{equation}
bound by the time range $t_1 \le t_p \le t_2$.
The subtraction of an event-unrelated background
flux $f_{\rm BG}$ is a particularly important 
correction for the smallest events of a size
distribution, when they become comparable or smaller
than the event-unrelated background flux. 

The fluences have the physical unit of [energy],
and are measured by the time integral of the 
time profile $f(t)$. Thus, the fluences $(E)$ are 
the time-integrated fluxes, 
\begin{equation}
	E=\int_{t_1}^{t_2} [f(t) - f_{\rm BG}] \ dt \ ,
\end{equation} 
for a time interval ($t_1,t_2$) with well-defined start
times $t_1$ and end times $t_2$). 

Details of the 
published power law slopes of size distributions
($\alpha_F$, $\alpha_E$) and references are given
in Aschwanden et al.~(2016); Aschwanden (2022, 2025), 
and Aschwanden and Gogus (2024), and are not
repeated here for brevity reasons. Most of the
processed information in this study is based on the
means $\mu$ and standard deviations $\sigma$ 
of the 23 astrophysical phenomena, for the 
SOC parameters $F$ and $E$, as listed in Table 1.

\subsection{	Non-Parametric Statistics	}

We tabulate our findings of power law slopes of
fluxes $\alpha_F$ and fluences $\alpha_E$ for
each of the astrophysical phenomena in Table 1
and show their histograms in Fig.~(4). 
The histograms in Fig.~(4) contain astrophysical
phenomena, which amounts to $N=15$ values for the
fluxes, and $N=12$ for the fluences, respectivly,
where we excluded $N=5$ non-SOC phenomena as 
described in the following. 

From the two samples of phenomena in the histograms 
shown in Fig.~(4) we calculate the means $\mu$ and
standard deviations $\sigma$ in the two histograms, 
which is a method of non-parametric statistics without 
any assumption on the functional shape of the 
size distributions $N(\alpha_F)$ and $N(\alpha_E)$.
We find a distribution of 
$\alpha_F=1.85\pm0.19$ for the fluxes, which is close
to the theoretical FD-SOC prediction $\alpha_{f,theo}=1.80$.
Similarly, we find a distribution of 
$\alpha_E=1.67\pm0.09$ for the fluences, which 
precisely matches the theoretical FD-SOC prediction 
$\alpha_{E,theo}=1.67$. Note that the uncertainty of the 
power law slopes varies in the range of 
$\sigma/\mu\approx 5\%-10\%$. 
Inspection of the deviation of power law slopes
from ideal straight power laws reveal that
most of the uncertainties occur for a variety of 
data analysis errors, such as
neglected background subtraction, inadequate 
fitting range, small-number statistics, or 
may naturally manifest a non-SOC process.

\subsection{	Power Law Slopes of Fluxes		}

Let us investigate the results of the power law slopes
$\alpha_F$ for fluxes, which are shown in form of histograms
$N(\alpha_F)$ in Fig.~4 (top panel), while the dashed vertical
line indicates the theoretically predicted value.
We find the following 15 parameters
with means $\mu$ and standard deviations $\sigma$:
$\alpha_{\rm HXR} =1.74\pm0.11$ for solar flare hard X-rays,
$\alpha_{\rm SXR} =1.87\pm0.10$ for solar flare soft X-rays,
$\alpha_{\rm EUV} =1.68\pm0.16$ for solar EUV nanoflares,
$\alpha_{\rm Inc} =1.80\pm0.21$ for solar incoherent radio bursts,
$\alpha_{\rm CME} =2.01\pm0.35$ for coronal mass ejections, 
$\alpha_{\rm Aur} =1.79\pm0.21$ for magnetospheric aurora events,
$\alpha_{\rm SF}  =1.82\pm0.37$ for stellar flares,
$\alpha_{\rm KEPLER}=1.87\pm0.29$ for stellar flares observed with KEPLER,
$\alpha_{\rm TESS}=2.27\pm0.19$ for stellar flares observed with TESS,
$\alpha_{\rm AGN} =1.73\pm0.19$ for active galactic nuclei events,
$\alpha_{\rm BH}  =1.65\pm0.25$ for black hole systems,
$\alpha_{\rm BL}  =1.98\pm0.08$ for blazars,
$\alpha_{\rm GRB} =2.10\pm0.25$ for gamma-ray bursts, and
$\alpha_{\rm SGR} =1.88\pm0.06$ for soft gamma-ray repeaters,
$\alpha_{\rm XB}  =1.52\pm0.45$ for X-ray binaries,
which are mostly consistent with the predictions of the
FD-SOC theoretical model within the statistical uncertainties, 
i.e., $\alpha_F=1.80$.
The variety of astrophysical source locations, ranging
from solar flares all the way to galactic events, 
underscores the universality of the FD-SOC model.
The most discrepant case is coherent radio emission,
($\alpha_{\rm Coh}=1.29\pm0.12$), while incoherent radio 
emission displays excellent agreement 
($\alpha_{\rm Inc}=1.80\pm0.21$) with the
FD-SOC model ($\alpha_F=1.80$). We will discuss below how
coherent and incoherent radio emission indicate different
plasma physics conditions.

\subsection{	Power Law Slopes of Fluences and Energies	}

The power law distributions for fluxes and fluences are
not identical, and hence we expect different slope values,
which are higher for fluxes, $\alpha_F=(9/5)=1.80$, and 
are lower for fluences or energies, $\alpha_E=(5/3)\approx 1.67$.
Therefore the FD-SOC model predicts this duality, which we
are testing here. In Fig.~4, the thin vertical line indicates
the observed mean power law slope $\mu$, while the dashed vertical
line depicts the theoretically predicted value.
From various astrophysical observations we find the 
following 12 parameters (Table 2):
$\alpha_{\rm HXR} =1.56\pm0.11$ for solar flare hard X-ray emission,
$\alpha_{\rm SXR} =1.79\pm0.35$ for solar flare soft X-ray emission,
$\alpha_{\rm CME} =1.84\pm0.41$ for coronal mass ejections, 
$\alpha_{\rm WIND}=1.70\pm0.17$ for solar wind fluctuations,
$\alpha_{\rm Aur} =1.60\pm0.13$ for magnetospheric aurora events,
$\alpha_{\rm TGF} =1.73\pm0.40$ for terrestrial gamma-ray flashes,
$\alpha_{\rm BH}  =1.73\pm0.25$ for black hole systems,
$\alpha_{\rm BL}  =1.63\pm0.17$ for blazars,
$\alpha_{\rm FRB} =1.67\pm0.12$ for fast radio burst events,
$\alpha_{\rm GRB} =1.68\pm0.24$ for gamma-ray bursts,
$\alpha_{\rm SGR} =1.68\pm0.12$ for soft gamma-ray repeaters, 
$\alpha_{\rm XB}  =1.53\pm0.04$ for X-ray binary events. 
The behavior of power law slopes is very similar for the
flux $\alpha_F$ (Fig.~4, top panel) and 
fluence $\alpha_E$ (Fig.~4, bottom panel). 
Most of the observed mean values $\mu$
are consistent with the theoretical predictions
of the FD-SOC model within the statistical uncertainties, 
except for 2 phenomena (SEP and cosmic rays) that are 
classified as non-SOC phenomena.

\subsection{	Statistical Outliers	}

The main hypothesis of this study is a statistical
test whether the power law slopes of observed flux and 
fluence size distributions ($\alpha_x^{obs}\pm\sigma_x$) are 
consistent with the theoretical FD-SOC model ($\alpha^{theo}$).
We found that 18 astrophysical phenomena are consistent
with the FD-SOC model within about one standard deviation 
(see Table 1), while 5 phenomena
are found not to be consistent (Table 2), as 
discussed in the following. 

The two phenomena of solar coherent radio bursts and
solar energetic particles (SEP) show relatively flat power 
law slopes of $\alpha_F \approx \alpha_E \approx 1.3\pm0.1$
that cannot be reproduced by a FD-SOC model (with $\gamma=1$), 
but can be produced by a generalized SOC model with a nonlinear
flux-volume coefficient of $\gamma \approx 2$.

Cosmic rays have a relatively steep power law slope of
$\alpha_E \approx 3.0\pm0.3$, which also cannot be produced
with the FD-SOC model, but matches the predicted slope
in a one-dimensional (1-D) Euclidean domain ($D_V=1$).

Random solar radio bursts follow the random distribution
of Poisson statistics by definition, which is contrary
to the nonlinear power law statistics of SOC systems
(with ``fat tails'' in their size distribution),
and thus consequently cannot be reproduced by the
FD-SOC model.

Finally, the fifth outlier, i.e., pulsar glitches,
exhibit erratic size distributions that cannot be
fitted with any power law distribution 
(see Cairns 2004; Melatos et al.~2008),
which probably results from small-number statistics, 
background subtraction problems, inadequate fitting
range, finite system-size effects, and other large 
deviations from ideal power laws.

In any case, what the 5 outliers have in common is
their incompatibility with the FD-SOC model, which
justifies their elimination from statistical tests, 
since they require different physical mechanisms. 
This is not a circular argument, because
the FD-SOC model predicts unique parameters
without free variables,
such as the peak flux slope $\alpha_F$ or the
fluence slope $\alpha_E$ of the investigated size
distributions. We do not aim to derive the functional
form of the statistical distribution $N(\alpha_x)$,
but merely perform a test whether the observed
power law slopes agree with the theoretically 
predicted values within the statistical uncertainties.

\subsection{	Solar Energetic Particle (SEP) Events	}

The anomaly of SEP events has been pointed out earlier and 
it was suggested that proton-emitting solar flares are a
special class of events, requiring a different physical
mechanism for the production of energetic protons 
(Cliver et al.~1991; Kahler 2013; Cliver and D'Huys 2018). 
In the derivation of the FD-SOC model we can distinguish
two different processes in the flare evolution, 
$F \propto V^\gamma$, with $\gamma=1$ for a slow linear
evolution, and with $\gamma \approx 2$ for a fast 
nonlinear evolution, which is also a measure of the 
spatio-temporal evolution $F(t) \propto V(t)^\gamma$ 
of an individual flare event, in the statistical average. 
The nonlinearity coefficient $\gamma$ is not
predicted by the FD-SOC model, but can be estimated from
the observations, by inverting the relationship
$\alpha_F=1 + (d-1)/(D_V \gamma)$, which yields
$\gamma=(d-1)/(\alpha_F-1)$. Inserting a typical
observational value of $\alpha_E \approx 1.4$, 
we obtain a coefficient of $\gamma=2$, for 
the Euclidean space dimension $d=3$. 
Thus, we can explain both the SEP events,
$\alpha_F=1.29\pm0.12$, as well as the
solar coherent radio burst events,
$\alpha_F=1.36\pm0.26$, with the generalized 
SOC model, matching a nonlinear flux-volume 
coefficient of $\gamma \approx 2$.

\subsection{	Incoherent and Coherent Emission	}

Why does coherent radio emission,
($\alpha_{\rm Coh}=1.29\pm0.12$), has a different size 
distribution than incoherent radio emission 
($\alpha_{\rm Inc}=1.80\pm0.21$)?
In the Introduction we pointed out that the standard
FD-SOC model is based on three fundamental assumptions
(fractality, diffusive transport, and flux-volume
proportionality), which defines a linear relationship
between the observed flux and (fractal) volume, i.e.,
$F \propto V^\gamma$ with $\gamma=1$ in the standard 
FD-SOC model. The resulting proportionality, 
$F \propto V$, implies then an equivalence between 
the corresponding power law slopes, i.e.,
$\alpha_F = \alpha_V$. This linear behavior is also
called an incoherent or a random process. 
Incoherent emission processes occurring in the solar 
corona include, for example, thermal bremsstrahlung,
gyroresonance emission, or gyro-synchrotron emission.  

In contrast, a nonlinear behavior is typical for coherent
emission, where the nonlinearity is expressed by the
relationship $F \propto V^\gamma$, for $\gamma > 1$.
Coherent emission entails exponential-growth processes
during an instability,
which is a typical nonlinear behavior. Such coherent emission
processes occur when a particle distribution function becomes
inverted by some dynamic process, such as by electron
beam formation, or by loss-cone distribution functions,
most conspicuously visible in electron-cyclotron microwave 
amplification by stimulated emission of radiation (MASER).
The phenomenon of electron beams is driven by a positive
gradient in the parallel velocity distribution function,
while loss-cones are driven by perpendicular positive
gradients in the velocity distribution function
(see textbooks on plasma physics, e.g., 
Benz 1993, Sturrock 1994, Boyd and Sanderson 2003).

\subsection{	Random Size Distributions		}

There is another anomaly of substantial deviations from
ideal power laws notable in at least 1 (out of the 23)
cases, which we call {\sl solar random radio bursts}
and has been reported in 4 cases: 
a type I storm with $\alpha_F=3.25\pm0.35$ 
(Mercier and Trottet 1997);
a DCIM-S radio spike burst event with $\alpha_F=2.99\pm0.63$;
(Aschwanden et al.~1998);
a microwave spikes event with $\alpha_F=7.40\pm0.40$
(Ning et al.~2007); and
a type I storm with $\alpha_F=4.80\pm0.10$
(Iwai et al.~(2013). 
This type of radio bursts clearly is not consistent
with neither the standard FD-SOC model ($\gamma=1$)
nor with the generalized FD-SOC model ($\gamma>1$). 
First of all,
the 4 observations reported here are outside the
theoretical physical range of SOC parameters, 
which is $1.0 \le \alpha_x \le 3.0$. 
Secondly, the power law fitting ranges are found to
extend over very small ranges (often less than a
decade), so that no reliable power law slope can
be fitted. The most likely explanation for the
too steep power law slopes is the confusion
between the power law inertial range and the
exponentially fall-off at the upper end of the
size distribution. This transition from a power
law to an exponential drop-off is dictated by 
the finite-system size limit of the largest events, 
which is expected to form a gradual roll-over
within the range of 
$3.0 \lapprox \alpha_x < \infty$.
 
Most generally,
the observed power law distribution functions
can be fitted with a three-part model that includes
(i) the flattening due to incomplete sampling of
small events, (ii) the initial range that can be
fitted with a pure power law function, and (iii)
the steepening due to finite-system size effects
for the largest events, approximated with an
exponential function, following Poisson statistics. 
Such a generalized
three-part size distribution function can be
described by (Aschwanden 2021),
\begin{equation}
        N(x) dx = N_0 (x_0 + x)^{-\alpha_x}
                  \exp{(-{x \over x_e})} \ dx \ ,
\end{equation}
Such a three-part size distribution may recover
the convolved power law slope $\alpha_x$,
but it may blurr the distinction between 
the FD-SOC model and a non-SOC model. 
The FD-SOC model requires a pure power law function,
while the non-SOC model presented here
requires a convolution of a power law with
an exponential function.
The inclusion of the exponential fall-off 
fits the Poisson statistics of
random processes.

\subsection{	Universal Constants			}

It is often said that power laws are the hallmarks
of SOC. Consequently, since power laws are measured 
by their slopes $\alpha_x$, we can also say that 
the (slopes of) size distributions are the hallmarks
of SOC. A variety of slope values $\alpha_x$ have been
reported within a range of $1.5 \le \alpha_x \le 2.3$.
The question arises now whether these constants
$\alpha_x$ are universally valid and predictable,
or are they individual for every astrophysical 
phenomenon and for every physical process, 
and then are likely to be unpredictable. Our FD-SOC model
suggests that every SOC parameter $x$ has a 
predictable power law slope $\alpha_x$ that
is universally valid, where the SOC parameters 
include the length scale $L$, the area $A$, 
the volume $V$, the flux $F$, the fluence or energy $E$,
and the duration $T$.
The predicted power law slope values are 
$\alpha_L={3}$,
$\alpha_A={7/3}$,
$\alpha_V={9/5}$,
$\alpha_F={9/5}$,
$\alpha_E={5/3}$, and 
$\alpha_T=2$.

Why are these SOC parameters universally valid?
The fundamental reason is that the underlying 
three assumptions are universal too, namely
the fractality, the diffusive transport, and
the flux-volume proportionality. The latter
assumption implies a constant emissivity, i.e.,
flux density $F$ per volume element $V$, 
i.e., $\varepsilon = F/V$. This assumption is 
an approximation only of course, but works
satisfactorily for a variety of incoherent 
emission mechanisms in terms of the electron
flux density or photon flux density (luminosity).
Hence, the universality of the power law slopes
$\alpha_x$ is a natural consquence of the
universal existence of the scaling laws
for the fractal dimension, $D_V=d-1/2$,
the diffusive transport, $L \propto T^{1/2}$,
and the flux-volume linearity, $F \propto V$, 
for a FD-SOC model with Euclidean 
dimensionality $d=3$.

\section{	SUMMARY AND CONCLUSIONS		}

Let us summarize our data analysis and conclusions:

\begin{enumerate}
\item{We extracted a total of $\sum N_F=170$ flux 
and $\sum N_E=104$ fluence
size distributions from the published literature, which are
characterized by the power law slopes $\alpha_F$ and $\alpha_E$.
We group these $274$ cases into 23 astrophysical phenomena,
ranging from solar, stellar, terrestrial, magnetospheric 
to galactic distances. Histograms of the power law slopes
are constructed for the fluxes and fluences separately.
Astrophysical phenomena with large deviations from ideal 
power laws are likely to be caused by background subtraction 
errors, inadequate fitting ranges, confusion with exponential 
fall-offs at the largest events, small-number statistics,
or may represent a non-SOC process altogether. 
The distinction between SOC processes and non-SOC processes 
is tabulated in the last column of Tables 1 and 2.}

\item{The means and standard deviations of the
observed power law slopes 
agrees well with the theoretial predictions
of the standard FD-SOC model 
(in 18 out of 23 astrophysical phenomena), with 
$\alpha_F=1.80$ for the flux slopes and  
$\alpha_E=1.67$ for the fluence slopes,
after elimination of non-SOC systems}.

\item{Five outliers of astrophysical 
phenomena are identified and interpreted
in terms of non-SOC systems, namely
coherent radio bursts ($\alpha_F=1.29\pm0.12$), 
solar energetic particle (SEP) events ($\alpha_F=1.36\pm0.26$
and $\alpha_E=1.34\pm0.15$), 
random radio bursts ($\alpha_F=4.80\pm0.10$), 
and pulsar glitches ($\alpha_E=1.90\pm0.78$), 
while all other cases are consistent with the standard
FD-SOC model. Coherent radio bursts include
type I and type III solar radio bursts and
can be fitted with a nonlinear flux-volume 
relationship, as expected for coherent radiation
mechanisms. 
Random radio bursts can be explained with the 
exponential fall-off sampled at the largest events 
of a size distribution 
according to Poisson statistics, 
which can assume any
steep power law slope of $\alpha_F \gapprox 3$.
The flatter power law slope observed in SEP events 
indicates that not all flares produce protons,
which appear to have a larger threshold in 
small solar flares. 
Pulsar glitches are clearly
non-SOC processes that cannot be fittted by
power law functions, given the observed large 
standard deviations ($\alpha_F=1.90\pm0.78$).
Hence, size distributions, their slopes $\alpha_x$,
and their outliers provide a valuable diagnostic
of erroneous power law fits and non-SOC processes.}
\end{enumerate}

Future efforts on self-organized criticality models may
focus on (i) impoved precursor background subtraction errors
(for instance when the flaring Sun eclipses the galactic 
center or other celestial X-ray sources),
(ii) inadequate fitting ranges (which occur in the presence
of strong deviations from ideal power law fits),
(iii) combined fitting of power law functions and exponential 
fall-offs near the largest events, and
(iv) small-number statistics (requiring larger data sets).
Further tasks are the calculation of realistic 
fluences and energies, especially the estimates and
frequency of the largest events that are essential
when extrapolating the SOC statistics of extreme
events. Other research subjects in SOC statistics are 
the extreme events of natural catastrophes, such as
earthquakes, forest fires, wild fires, mountain slides, 
mud slides, hurricanes, taifuns, global climate changes,
epidemics and pandemics (such as Covid-19),
for which SOC models all have been found to be relevant.

\acknowledgments
{\sl Acknowledgements:}
We acknowledge constructive and stimulating discussions
(in alphabetical order)
with Arnold Benz, Sandra Chapman, Paul Charbonneau, 
Henrik Jeldtoft Jensen, Adam Kowalski, Sam Krucker, 
Alexander Milovanov, Leonty Miroshnichenko, Jens Juul 
Rasmussen, Karel Schrijver, Vadim Uritsky, Loukas Vlahos, 
and Nick Watkins.
This work was partially supported by NASA contract NNX11A099G
``Self-organized criticality in solar physics'' and NASA contract
NNG04EA00C of the SDO/AIA instrument to LMSAL.


\section*{      References      }
\def\ref#1{\par\noindent\hangindent1cm {#1}}

\ref{Aschwanden, M.J., Dennis, B.R., and Benz, A.O. 1998,
        {\sl Logistic avalanche processes, elementary time structures,
        and frequency distributions of flares},
        ApJ 497, 972.}
\ref{Aschwanden, M.J. 2011a,
        {\sl Self-Organized Criticality in Astrophysics. The Statistics
        of Nonlinear Processes in the Universe},
        Springer-Praxis: New York, 416p.}
\ref{Aschwanden, M.J. 2014,
        {\sl A macroscopic description of self-organized systems and
        astrophysical applications}, ApJ 782, 54.}
\ref{Aschwanden, M.J. 2015,
        {\sl Thresholded power law size distributions of instabilities
        in astrophysics}, ApJ 814.}
\ref{Aschwanden, M.J. 2013a, in {\sl Theoretical Models of SOC Systems},
        chapter 2 in {\sl Self-Organized Criticality Systems}
        (ed. Aschwanden M.J.), Open Academic Press: Berlin, Warsaw,
        www.openacademicpress.de p.21.}
\ref{Aschwanden,M.J. 2013b,
        {\sl Self-Organized Criticality Systems in Astrophysics (Chapter 13)},
        in "Self-Organized Criticality Systems" (ed. Aschwanden,M.J.),
        Open Academic Press: Berlin, Warsaw, p.439.}
\ref{Aschwanden, M.J., Crosby, N., Dimitropoulou, M., Georgoulis, M.K.,
        Hergarten, S., McAteer, J., Milovanov, A., Mineshige, S.,
        Morales, L., Nishizuka, N., Pruessner, G., Sanchez, R.,
        Sharma, S., Strugarek, A., and Uritsky, V. 2016,
        {\sl 25 Years of Self-Organized Criticality: Solar and
        Astrophysics}, Space Science Reviews 198, 47.}
\ref{Aschwanden,M.J. 2019a,
        {\sl New millennium solar physics},
        Springer Nature: Switzerland, Science Library 458.}
\ref{Aschwanden, M.J. 2021,
        {\sl Finite system-size effects in self-organizing criticality 
	systems}, ApJ 909.}
\ref{Aschwanden, M.J. 2022,
        {\sl The fractality and size distributions of astrophysical
        self-organized criticality systems},
        ApJ 934 33}
\ref{Aschwanden, M.J. and Gogus, E. 2024,
	{\sl Testing the universality of self-organized criticality
	in galactic, extra-galactic, and black-hole systems},
	ApJ (in press).}
\ref{Aschwanden, M.J. 2025,
	{\sl Power Laws in Astrophysics. Self-Organzed Criticality
	Systems}, Cambridge University Press: Cambridge.}
\ref{Bak, P., Tang, C., and Wiesenfeld, K. 1987,
        {\sl Self-organized criticality: An explanation of 1/f noise},
        Physical Review Lett. 59(27), 381.}
\ref{Bak, P. 1996,
        {\sl How Nature Works. The Science of Self-Organized Criticality},
        Copernicus: New York.}
\ref{Benz, A.O. 1993, {\sl Plasma Astrophysics},
 	Kluwer Academic Publishers: Dordrecht, The Netherlands.}
\ref{Boyd, T.J.M. and Sanderson,J.J. 2003, {\sl The Physics of
	Plasmas}, Cambridge University Press: Cambridge.}
\ref{Cairns, I.H. 2004,
        {Properties and interpretations of giant micropulses
        and giant pulses from pulsars}, ApJ 610, 948-955.}
\ref{Charbonneau,P. 2017, 
        {\sl Natural complexity: A modeling handbook},
        Princeton University Press, Princeton, New Jersey.}
\ref{Cliver, E., Reames, D., Kahler, S., and Cane, H. 1991,
        {\sl Size distribution of solar energetic particle events},
        Internat. Cosmic Ray Conf. 22nd, Dublin, LEAC A92-36806 15-93,
        NASA:Greenbelt, p.2:1.}
\ref{Cliver, E. and D'Huys, E. 2018,
        {\sl Size distributions of solar proton events and their associated
        soft X-ray flares},
        ApJ 864, 48/1, 11.}
\ref{Galam, S. 2012, {\sl Sociophysics. A physicist's modeling of
        psycho-political phenomena}, Berlin: Springer.}
\ref{Jensen, H.J. 1998, {\sl Self-Organized Criticality. Emergent
        Complex Behavior in Physical and Biological Systems},
        Cambridge University Press: Cambridge.}
\ref{Ilachinski,A. 2001,
        {\sl Cellular automata},
        World Scientific: New Jersey, 840p.}
\ref{Iwai,K., Masuda, S., Miyoshi, Y., Tsuchiya, F., Morioka, A., and
        Misawa, H. 2013, {\sl Peak Flux Distributions of Solar Radio Type-I
        Bursts from Highly Resolved Spectral Observations},
        ApJ 768, L2.}
\ref{Hergarten, S. 2002, {\sl Self-Organized Criticality in Earth systems},
	Springer: Berlin.}t
\ref{Kahler, S.W. 2013,
        {\sl Does a Scaling Law Exist between Solar Energetic Particle
        Events and Solar Flares?}, ApJ 769, 35.}
\ref{McAteer,R.T.J., Aschwanden,M.J., Dimitropoulou,M., Georgoulis,M.K.,
        Pruessner, G., Morales, L., Ireland, J., and Abramenko,V. 2016,
        {\sl 25 Years of Self-Organized Criticality: Numerical Detection 
	Methods}, SSRv 198, 217.}
\ref{Melatos, A., Peralta, C., and Wyithe, J.S.B. 2008,
        {\sl Avalanche Dynamics of radio pulsar glitches},
        ApJ 672, 1103-1118.}
\ref{Mercier, C. and Trottet, G. 1997,
        {\sl Coronal radio bursts: A signature of nanoflares ?},
        ApJ 484, L65.}
\ref{Ning, Z., Wu,H., and Meng, X. 2007, {\sl Frequency distributions
	of microwave pulses for the 18 March 2003 solar flare},
	Sol.Phys. 242, 101.}
\ref{Pruessner, G. 2012, {\sl Self-Organised Criticality. Theory, Models
        and Characterisation}, Cambridge University Press: Cambridge.}
\ref{Scott,A.C. 2007,
        {\sl The nonlinear universe: chaos, emergence, life},
        Springer: Berlin, 364 p.}
\ref{Sharma,A.S., Aschwanden,M.J., Crosby,N.B., Klimas,A.J., Milovanov,A.V.,
        Morales,L., Sanchez,R., and Uritsky,V. 2016,
        {\sl 25 Years of Self-Organized Criticality: Space and Laboratory 
	Plamsas}, SSRv 198, 167.}
\ref{Sornette, D. 2004,
        {\sl Critical phenomena in natural sciences: chaos, fractals,
        self-organi\-za\-tion and disorder: concepts and tools},
        Springer, Heidelberg, 528 p.}
\ref{Sturrock, P.A. 1994, {\sl Plasma Physics},
 	Cambridge University Press: Cambridge}
\ref{Uritsky, V.M., Klimas, A.J., Vassiliadis, D., Chua, D., and
	George Parks 2002, {\sl Scale-free statistics of spatiotemporal
	auroral emissions as depicted by POLAR UVI images: Dynamic
	magnetosphere is an avalanchin system},
	JGR 107, A12, 1426,}
\ref{Watkins, N.W., Pruessner, G., Chapman, S.C., Crosby, N.B., and Jensen, H.J.
        {\sl 25 Years of Self-organized Criticality: Concepts and 
	Controversies}, 2016, SSRv 198, 3.}

\clearpage
\begin{table}
\begin{center}
\caption{Astrophysical phenomena (column 1), 
number of flux data sets $N_F$ (column 2),
the power law slope of flux distributions $\alpha_F$ (column 3), 
the number of fluence data sets $N_E$ (column 4),
the power law slope of fluence or energy $\alpha_E$ (column 5),
and the interpretation of SOC and non-SOC models (columm 6).}

\normalsize
\medskip
\begin{tabular}{llllll}
\hline
Astrophysical                 &         & Power         &           & Power      & Theoretical \\
phenomena                     &         & law           &           & law        & Interpretation \\
                              &         & slope         &           & slope      & \\
                              & $N_F$   & $\alpha_F$   & $N_E$      & $\alpha_E$ & \\
\hline
\hline
Solar Flare Hard X-Rays (HXR) & 20      & 1.74$\pm$0.11 & 9         & 1.56$\pm$0.11  & SOC\\
Solar Flare Soft X-Rays (SXR) & 10      & 1.87$\pm$0.10 & 5         & 1.79$\pm$0.35  & SOC\\
Solar Nanoflares (EUV)        & 12      & 1.68$\pm$0.16 & 0         & ...            & SOC\\
Solar Incoherent Radio Bursts & 7       & 1.80$\pm$0.21 & 0         & ...            & SOC\\
Solar Coronal Mass Ejections (CME) & 5  & 2.01$\pm$0.35 & 6         & 1.84$\pm$0.41  & SOC\\
Solar Wind (WIND)             & 3       & ...           & 13        & 1.70$\pm$0.17  & SOC\\
Magnetospheric Auroras        & 12      & 1.79$\pm$0.21 & 10        & 1.60$\pm$0.13  & SOC\\
Terrestrial Gamma-Ray Flashes (TGF)& 0  & ...           & 3         & 1.73$\pm$0.40  & SOC\\
Stellar Flares                & 15      & 1.82$\pm$0.37 & 0         & ...            & SOC\\
Stellar Flares KEPLER         & 44      & 1.87$\pm$0.29 & 0         & ...            & SOC\\
Stellar Flares (TESS)         & 5       & 2.27$\pm$0.19 & 0         & ...            & SOC\\
Active Galactic Nuclei (AGN)  & 2       & 1.73$\pm$0.19 & 0         & ...            & SOC\\
Black-Hole Systems (BH)       & 1       & 1.65$\pm$0.17 & 1         & 1.73$\pm$0.25  & SOC\\
Blazars (BL)                  & 2       & 1.98$\pm$0.08 & 2         & 1.63$\pm$0.17  & SOC\\
Fast Radio Bursts (FRB)       & 0       & ...           & 13        & 1.67$\pm$0.12  & SOC\\
Gamma Ray Bursts (GRB)        & 5       & 2.10$\pm$0.25 & 5         & 1.68$\pm$0.24  & SOC\\
Soft Gamma Ray Repeaters (SGR)& 2       & 1.88$\pm$0.06 & 18        & 1.68$\pm$0.12  & SOC\\
X-Ray Binaries (XB)           & 6       & 1.52$\pm$0.45 & 4         & 1.53$\pm$0.04  & SOC\\
\hline
Observations Means            & 15      & 1.85$\pm$0.19 & 12        & 1.67$\pm$0.09  & SOC\\
FD-SOC Prediction             &         & {\bf 1.80}    &           & {\bf 1.67}     & SOC\\
\hline 
\end{tabular}
\end{center}
\end{table}

\begin{table}
\begin{center}
\caption{Statistical outliers of astrophysical phenomena (see discussion in Section 2.6).} 
\normalsize
\medskip
\begin{tabular}{llllll}
\hline
Astrophysical                 &         & Power         &           & Power      & Theoretical \\
phenomena                     &         & law           &           & law        & Interpretation \\
                              &         & slope         &           & slope      & \\
                              & $N_F$   & $\alpha_F$   & $N_E$      & $\alpha_E$ & \\
\hline 
\hline
Solar Coherent Radio Bursts   & 5       &(1.29$\pm$0.12)& 0         & ...            & gen-SOC, $\gamma\approx 2$\\
Solar Energetic Particles (SEP)& 7      &(1.36$\pm$0.26)& 12        &(1.34$\pm$0.15) & gen-SOC, $\gamma\approx 2$\\
Cosmic Rays                   & 0       &               & 3         &(3.02$\pm$0.03) & gen-SOC, $D_V=1.0$\\
Solar Random Radio Bursts     & 4       &(4.80$\pm$0.10)& 0         & ...            & non-SOC, random\\
Pulsar Glitches               & 3       &(1.90$\pm$0.78)& 0         & ...            & non-SOC, erratic\\
\hline 
\end{tabular}
\end{center}
\end{table}
\clearpage

\begin{figure}[t]
 \centerline{\includegraphics[width=0.4\textwidth]{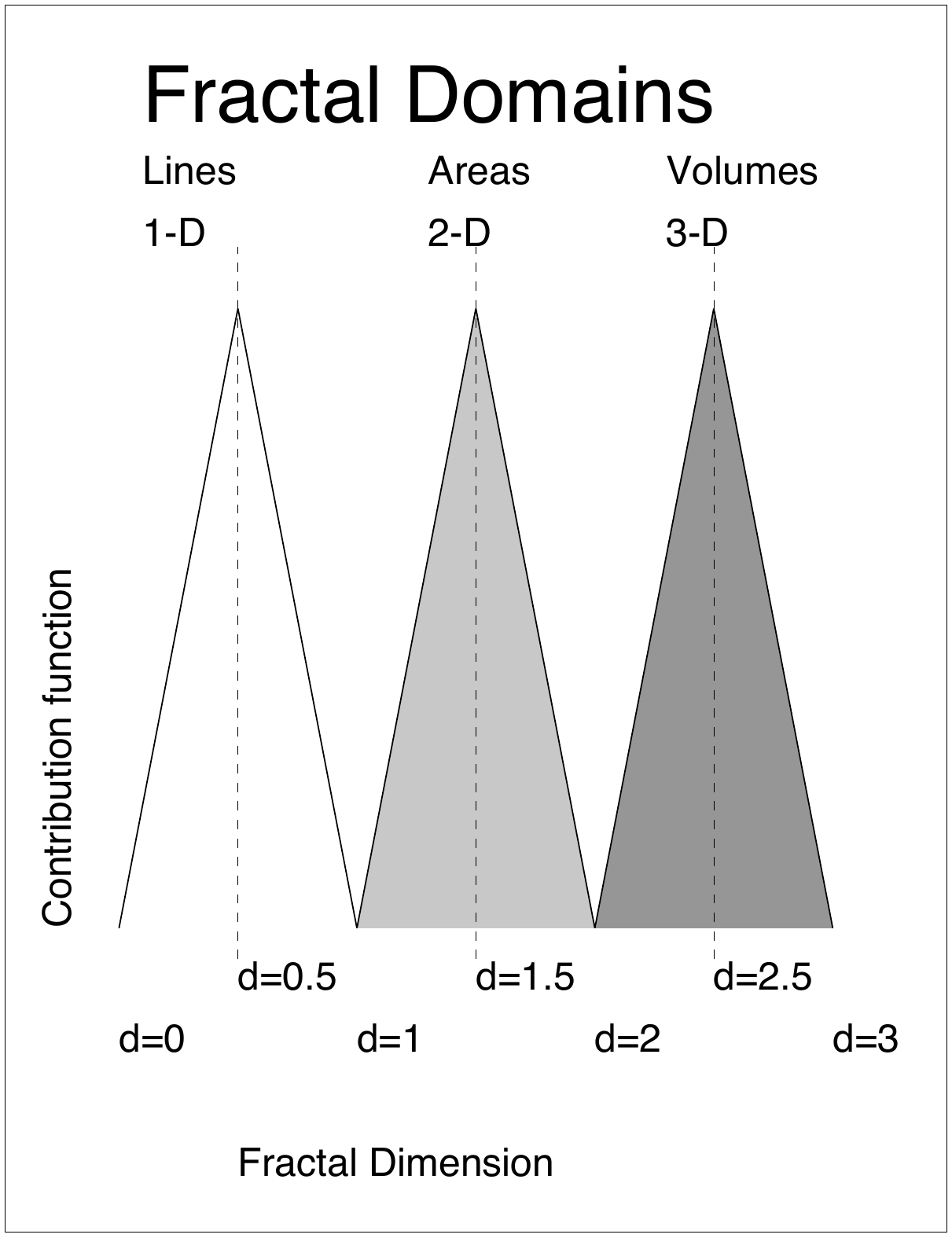}}
\caption{The contribution function to the three fractal domains
(1-D lines, 2-D areas, and 3-D volumes is illustrated as a function 
of the fractal dimension $0 \le d \le 3$. The mean fractal dimensions
are $d$=0.5, 1.5, and 2.5 in these 3 domains.}
\end{figure}

\begin{figure}[h]
 \centerline{\includegraphics[width=0.9\textwidth]{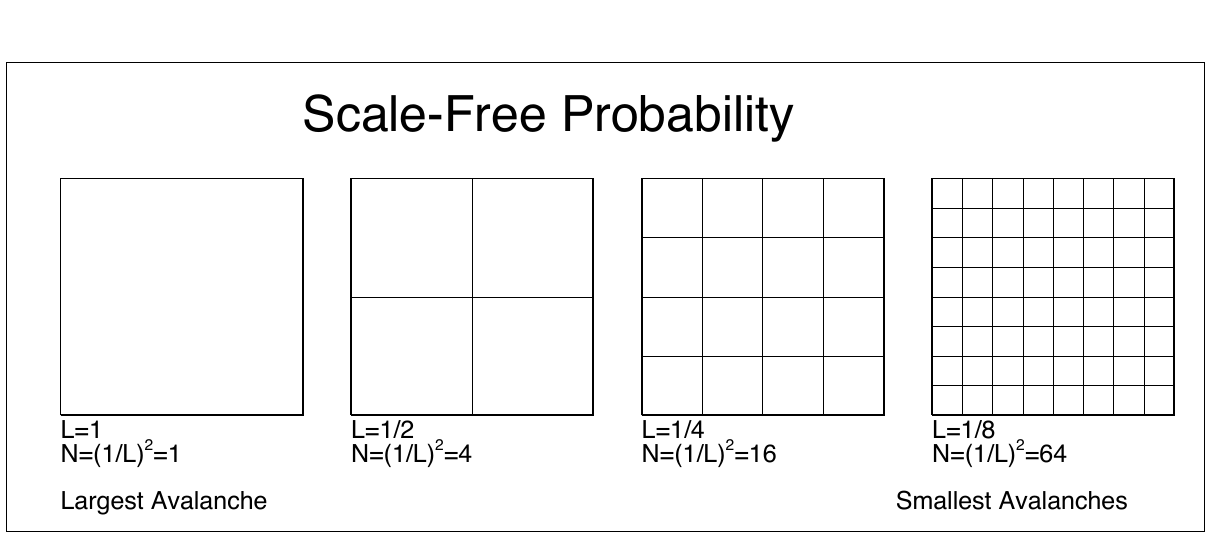}}
\caption{The scale-free probability is illustrated by segmenting
the largest avalanche area (on the left side) into smaller avalanche
areas (on the right side). While the length scale $L$ decreases by 
a factor of 2 in each frame, the probability increases reciprocally 
by a factor of $N=2^d=4$ for the Euclidean dimension of $d=2$, 
obeying the reciprocal scaling law $N=L^{-d}$.}
\end{figure}

\begin{figure}[t]
 \centerline{\includegraphics[width=0.5\textwidth]{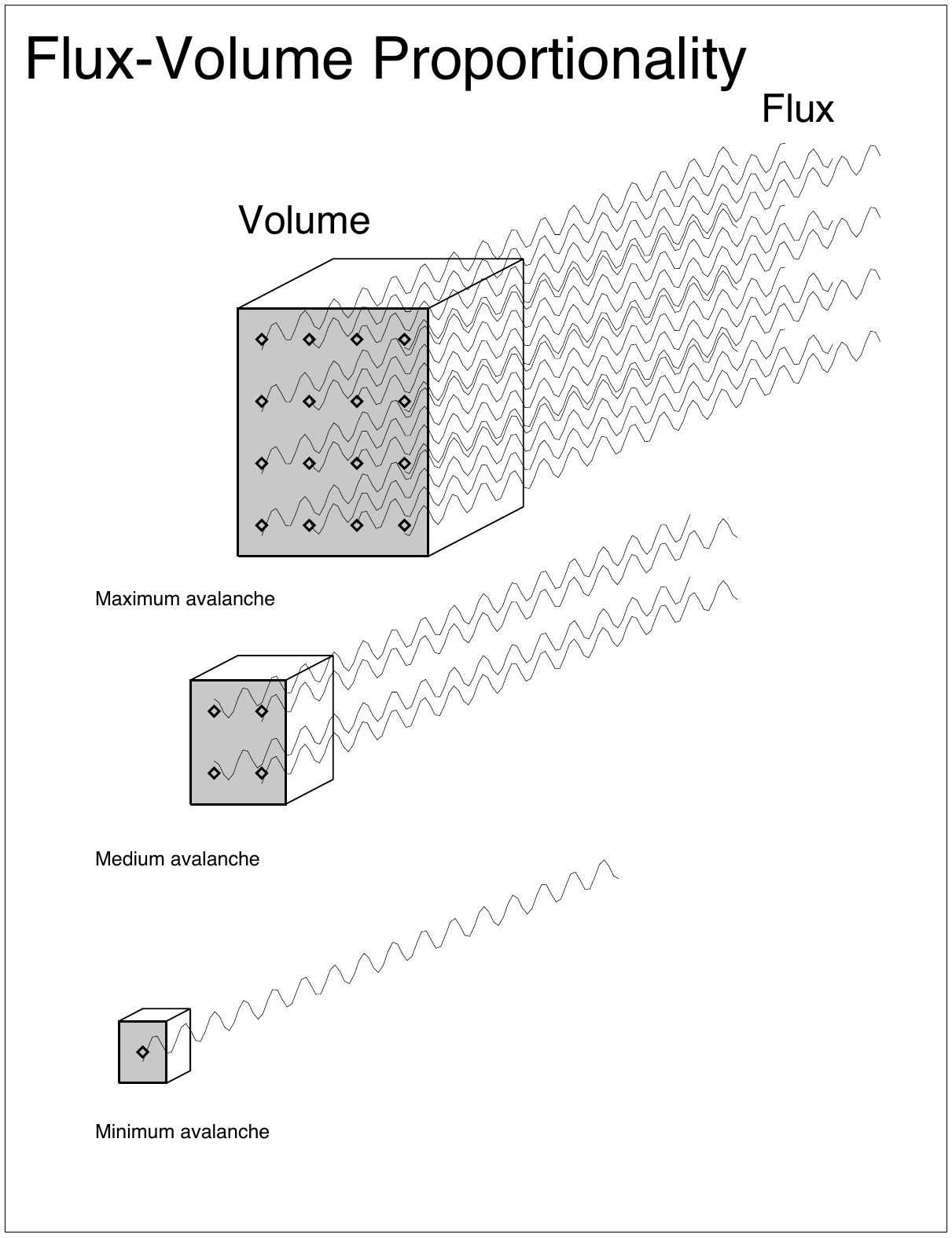}}
\caption{The relationship between the fractal volume $V$ and the observed 
flux $F$ is depicted for three different avalanche sizes, $F \propto V$.}
\end{figure}

\begin{figure}
 \centerline{\includegraphics[width=0.8\textwidth]{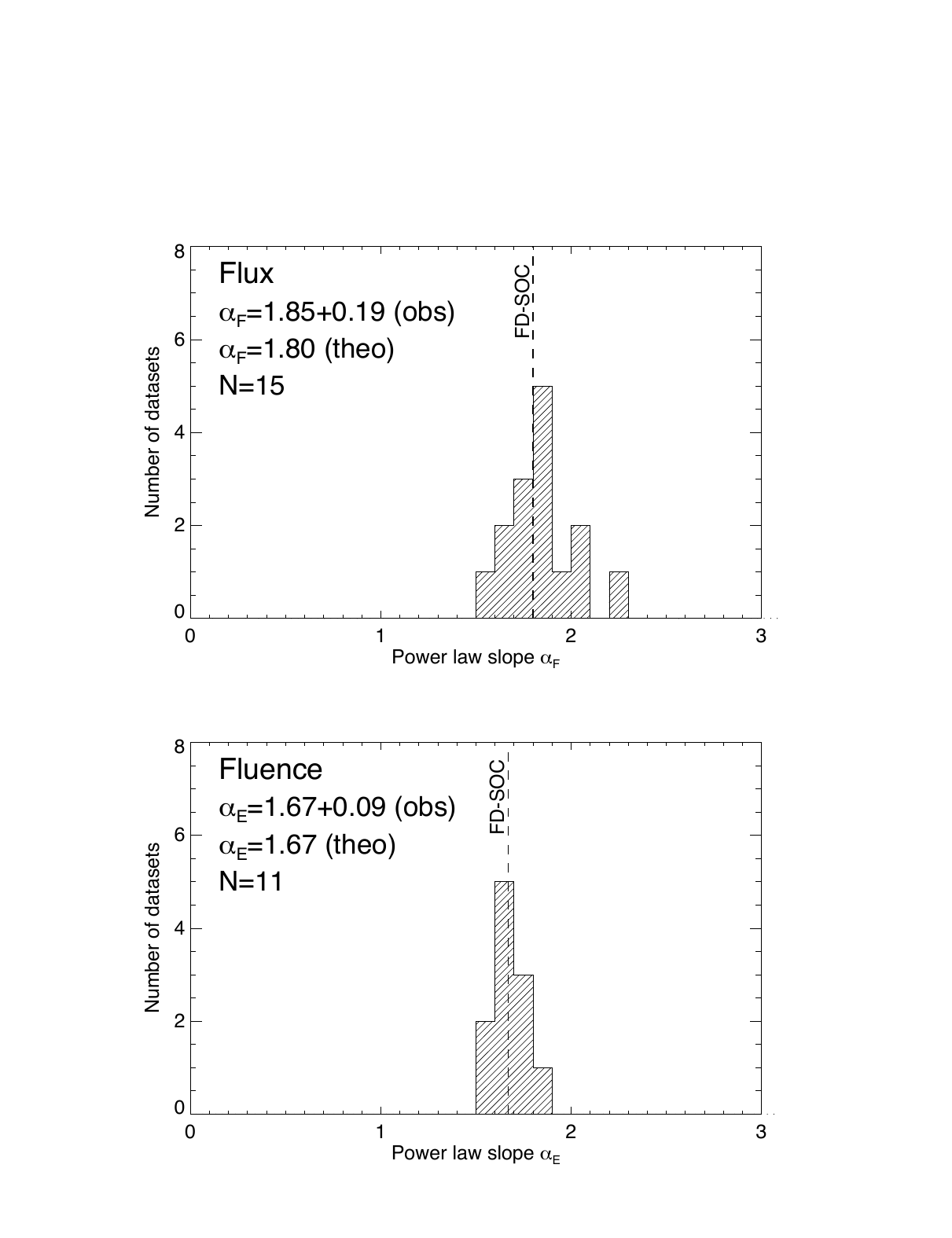}}
\caption{Histograms of 
power law slopes $\alpha_F$ for fluxes $F$ (top panel) and of 
power law slopes $\alpha_E$ for fluences $E$ (bottom panel),
where each statistical element corresponds to a different
astrophysical phenomenon, as tabulated in Table 1.
The number of phenomena is indicated with the symbol $N$,
and the the theoretical FD-SOC prediction
is indicated with a vertical dashed line.}
\end{figure}

\end{document}